\documentstyle[12pt,epsf,epsfig]{article}
\newcommand{\uq}{{\sf u}}                  % u quark
\newcommand{\dq}{{\sf d}}                  % d quark
\newcommand{\sq}{{\sf s}}                  % s quark
\newcommand{\cq}{{\sf c}}                  % c quark
\newcommand{\bq}{{\sf b}}                  % b quark
\newcommand{\qq}{{\sf q}}                  % quark
\newcommand{\QQ}{{\sf Q}}                  % heavy quark
\begin{document}
%\baselineskip=24truept

\date{\normalsize April 1997}

\title{\Large {\bf POLDIS} \\
A Monte Carlo for POLarized (semi-inclusive) \\
Deep Inelastic Scattering}

\author{Alessandro Bravar,$^{1,\ast}$ Krzysztof Kurek,$^2$
and Roland Windmolders~$^3$ \\
\\
\normalsize $^1$ Institut f\"ur Kernphysik, Universit\"at Mainz,
D-55099 Mainz, Germany \\
\normalsize $^2$ Soltan Institute for Nuclear Studies
and Warsaw University, 00681 Warsaw, Poland \\
\normalsize $^3$ Universit\'e de Mons-Hainaut, B-7000 Mons, Belgium \\}

\maketitle

\begin{abstract}
%\baselineskip=24truept
{\tt POLDIS} is a Monte Carlo program for polarized (semi-inclusive)
deep inelastic scattering (DIS).
Unpolarized DIS events are generated with the existing
lepto-production event generators {\tt LEPTO} for DIS 
and {\tt AROMA} for Heavy Flavor production.
The relevant spin asymmetries are computed at partonic level
to first order in $\alpha_s$ for each generated event,
and are then convoluted with the corresponding ratio
between the polarized and unpolarized parton distribution functions
({\it i.e.} parton polarization).
This procedure provides a {\it polarization weight} for each event.
The average of these {\it polarization weights} gives
the polarized cross section spin-asymmetry for the generated sample.
The code consists of a set of subroutines to be linked with
{\tt LEPTO} and/or {\tt AROMA}.
No modification to these programs is required.
Some existing polarized parton distribution functions
are also included. 
\end{abstract}

\vspace*{\fill}
\begin{center}
{\it To be submitted to Computer Physics Communications}
\end{center}

\noindent $^\ast$ Corresponding author: bravar@cern.ch

%%%%%%%%%%
\newpage
%\begin{center}
%{\large \bf LONG WRITE-UP}
%\end{center}
%\vspace*{-12pt}
%%%%%%%%%%
\section{
\label{sec:introd}
Introduction}

Polarized lepton-nucleon deep inelastic scattering (DIS) has 
been studied in the last decades by several experiments
which have measured spin asymmetries over a wide kinematic
range~\cite{Ash89,Sti96}.
These experiments have determined the spin structure functions
of the proton and the neutron and have tested the related sum rules.
When interpreted in the framework of the quark-parton model the experimental 
results show that the quark spins account for only a rather small fraction
of the nucleon spin,
thus implying an appreciable contribution either of gluons
or possibly of orbital angular momentum.
These data indicate also a large positive contribution of \uq~quarks,
a negative contribution of \dq~quarks,
and, surprisingly, a small negative contribution of
\sq~quarks to the proton spin~\cite{Ash89}.
A general introduction to this subject can be found for example
in~\cite{Ans95,Lea96}.

Inclusive polarized DIS measurements do not allow one to distinguish
the role of each individual partonic component.
A further separation of the contributions of different constituents
to the nucleon spin, like $\Delta {\sf s}$ or $\Delta G$,
requires additional input from the study of semi-inclusive DIS,
for which only limited data have been obtained so far~\cite{Ade96}.
In these experiments, in addition to the scattered lepton,
one detects also one or more hadrons produced in the interaction.
For instance, the study of polarized open-charm lepto-production
allows to access the gluon polarization $\Delta G$ in a polarized
nucleon~\cite{comp}.

In this work we present {\tt POLDIS}, a program designed to simulate
polarized DIS experiments, with particular emphasis on semi-inclusive DIS 
and Heavy Flavor lepto-production in the quasi-real photo-production
limit ($Q^2 \rightarrow 0$).
For these processes, {\tt POLDIS} generates the spin-dependent cross section
asymmetries between parallel and antiparallel configurations of
the incident lepton beam and target nucleon (or proton beam) polarizations,
which are measured in these experiments.
Since these asymmetries are ratios of cross sections
no absolute normalization is needed in their evaluation.
The spin-dependent cross sections can be extracted from these spin asymmetries
and the spin-independent cross section.

In this program electromagnetic processes mediated by one photon exchange
are implemented.
The present code can be used over a wide kinematical range,
where the effects of the weak interaction can be neglected,
{\it i.e.} for $Q^2 \leq 100~{\rm GeV}^2 \ll M_{Z^0}^2$.

The implementation of the polarization in {\tt POLDIS}
can be summarized in the following steps:
\begin{itemize}
\item[1 --] generation of an unpolarized event,
\item[2 --] calculation of the partonic level hard-scattering spin asymmetry
for this event,
\item[3 --] evaluation of the final spin asymmetry and of the spin-dependent
cross sections.
\end{itemize}

The unpolarized event generation is performed with the 
{\tt LEPTO}~\cite{lepto} Monte Carlo and the {\tt AROMA}~\cite{aroma} code
for Heavy Flavor (HF) production.
The hadronization is based on the LUND string model,
which is known to reproduce fairly accurately the final hadronic state
in a variety of processes, and is performed with {\tt JETSET}~\cite{jetset}.
The hard-scattering spin asymmetries are calculated for each generated event
to order $\alpha_s$\footnote{At present, the spin dependent hard
cross sections are calculated to order $\alpha_s$ only.}
and are convoluted with the ratio between the corresponding polarized 
and unpolarized parton densities ({\it i.e.} parton polarization).
A {\it polarization asymmetry weight} is thus obtained,
and the average of these {\it weights} for the generated sample
gives the polarized cross section asymmetry.
The spin-dependent cross sections can be obtained from this asymmetry
and the spin-independent cross section.
These calculations are performed in a set of subroutines to be linked
with the existing unpolarized lepto-production event generator
{\tt LEPTO}~\cite{lepto} for DIS
and {\tt AROMA}~\cite{aroma} for HF production,
and {\tt JETSET}~\cite{jetset} for the hadronization.
No modification to these programs is required.
The unpolarized parton densities are obtained from the {\tt PDFLIB}
library~\cite{pdflib};
various polarized parton densities can be selected from a collection of some existing parametrizations provided with this program,
although in a less standardized form.
{\tt POLDIS} has been tested with the most recent versions of these programs
({\tt LEPTO 6.5}, {\tt AROMA 2.2}, and {\tt JETSET 7.4});
it is also backward compatible with older versions.

A Monte Carlo code for polarized DIS with similar aims, {\tt PEPSI},
has been presented a few years ago~\cite{pepsi}.
Similar results could, in principle, be obtained with {\tt PEPSI};
however no Heavy Flavor generation is included in that program.
In order to generate the spin asymmetries, 
which are indeed the measured quantities,
it requires separate runs for opposite spin configuration
in addition to a run without polarization.
This results in a less convenient usage
compared to that adopted in {\tt POLDIS}.
Additionally, {\tt POLDIS} can also generate simultaneously different
asymmetry values using different polarized parton densities.

In the next Section we present the kinematics, formalism, and formulae
for the polarized DIS.
The partonic level hard-scattering spin-independent and spin-dependent 
cross sections, calculated to first order in $\alpha_s$
are summarized in Appendix~A and~B.
Section~3 describes the structure of the program and the implementation
of the physics presented in Section~2.
The usage of the program is explained in Section~4.
We conclude in Section~5 with a comparison between the scattering
asymmetries simulated by our program for some reactions
and experimental data.

%%%%%%%%%%
\section{
\label{sec:asym}
Polarized cross sections}

%%%%%%%%%%
\subsection{
\label{sec:kinema}
Kinematics}

\begin{figure}
\vspace*{-10mm}
\begin{center}
\mbox{\epsfxsize=12cm\epsffile{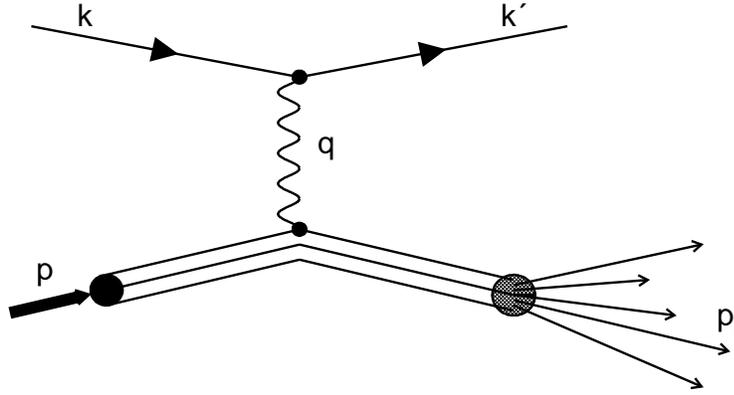}}
\end{center}
\vspace*{-5mm}
\caption{Deep inelastic scattering event.}
\label{fig:kinema}
\end{figure}

Figure~\ref{fig:kinema} depicts a deep inelastic scattering (DIS) event.
The four-vectors $k^\mu = (E,\vec{k})$ and 
$k^{\prime\mu} = (E^\prime,\vec{k}^\prime)$ represent
the momenta of the incoming and scattered lepton, respectively,
and $q^\mu = k^\mu - k^{\prime\mu}$ is the momentum transfer
from the lepton to the hadron ($\gamma^\ast$ four-momentum).
The target nucleon of mass $M$ has four-momentum $p^\mu$,
and $p^\mu_i$ is the four-momentum of the $i^{th}$ hadron produced in the
interaction.
The interaction is usually described with the following variables:
%%%%%
\begin{equation}
\begin{array}{lclcl}
Q^2 & = & - q^2             & \approx & 2 E E^\prime (1 - \cos \theta) \\ [6pt]
\nu & = & p \cdot q / M           & = & E - E^\prime \\ [6pt]
  y & = & p \cdot q / p \cdot k   & = & \nu / E \\ [6pt]
  x & = & Q^2 / 2 p \cdot q       & = & Q^2 / 2 M \nu \; .
\end{array}
\label{eq:kinema}
\end{equation}
%%%%%
The right-hand side of each equation is valid only in the laboratory frame,
where the target nucleon is at rest, and
$\theta$ is the lepton scattering angle.
The kinematics for inclusive scattering, integrated in azimuth,
is completely described by two of the four variables given above,
say $x$ and $Q^2$.
When describing semi-inclusive scattering, three additional variables
are needed for each measured hadron;
a common choice for these variables is
the energy fraction of the hadron with respect to the $\gamma^\ast$ energy 
%%%%%
\begin{equation}
z_i = p_i \cdot p / p \cdot q = E_i / \nu \; ,
\label{eq:sikinema}
\end{equation}
%%%%%
the hadron transverse momentum $p_T$ with respect to the $\gamma^\ast$ 
direction, and the azimuthal angle between the scattered lepton and
the outgoing hadron.
 
%%%%%
%%%%%

%%%%%%%%%%
\subsection{
\label{sec:xsecasymm}
Cross section asymmetries}

In a polarized DIS experiment one measures the asymmetries
\begin{equation}
A_\parallel = \frac{d\sigma^{\uparrow\downarrow} - \,
               d\sigma^{\uparrow\uparrow}}
              {d\sigma^{\uparrow\downarrow} + \,
               d\sigma^{\uparrow\uparrow}}
\; \; \; \; \; {\rm and} \; \; \; \; \; 
A_\perp = \frac{d\sigma^{\downarrow\rightarrow} - \,
               d\sigma^{\uparrow\rightarrow}}
              {d\sigma^{\downarrow\rightarrow} + \,
               d\sigma^{\uparrow\rightarrow}}
\label{eq:asym1}
\end{equation}
for longitudinal and transverse configurations of the incident
lepton and target polarizations.
The spin orientations in Eq.~\ref{eq:asym1} refer to the laboratory frame, 
where the target nucleon is at rest.
These asymmetries are directly related to the polarized structure functions
$g_1$ and $g_2$.
In this paper only the longitudinal asymmetry
$A_\parallel$ is discussed (and included in {\tt POLDIS}).
In the next pages, we will use the following notation for this asymmetry:
$A_{LL} \equiv A_\parallel$.

Usually the scattering asymmetry results are presented in terms of
the virtual photon asymmetries $A_1$ and $A_2$
%%%%%
\begin{equation}
A_1 = \frac{\sigma^T_{1/2}-\sigma^T_{3/2}}{\sigma^T_{1/2}+\sigma^T_{3/2}}
%= \frac{g_1-\gamma^2 g_2}{F_1} 
\; \; \; \; \; \;
{\rm and} \; \; \; \; \; \;
A_2 = \frac{2 \sigma^{TL}}{\sigma^T_{1/2}+\sigma^T_{3/2}}
\label{eq:asym2}
\end{equation}
%%%%%
where $\sigma^T_J$ is the virtual photon absorption cross section
in a configuration with total angular momentum $J$ along the
incident photon direction,
and $\sigma^{TL}$ is the interference term between transverse and longitudinal
virtual photon nucleon scattering.

$A_1$ and $A_2$ are related to the measured asymmetry $A_{LL}$ by
%%%%%
\begin{equation}
A_{LL} = D \, (A_1 + \eta A_2)
\label{eq:asym3}
\end{equation}
%%%%%
where
%%%%%
\begin{equation}
D \approx \frac{y(y-2)}{y^2 + 2(1-y)(1+R)} \; , \; \; \; \; \; \; \; \; \; \;
\eta \approx \frac{2(1-y)}{y(2-y)} \frac{\sqrt{Q^2}}{E}
\label{eq:depol}
\end{equation}
%%%%%
in the high energy limit (large $\nu$) and neglecting
the incident lepton mass.
$D$ is the depolarization factor of the virtual photon with respect to
the incident lepton, and 
%%%%%
\begin{equation}
R = \frac{\sigma^L}{\sigma^T}
\label{eq:rpar}
\end{equation}
%%%%%
is the ratio between the unpolarized cross section for the longitudinal
and transverse virtual photon components.
The ratio $R = R(x,Q^2)$ can be obtained from the QCD analysis of
unpolarized inclusive DIS data.
In practice, however, one uses parametrizations of $R$ obtained directly from
DIS experiments (see Section~\ref{sec:rpar}).

The virtual photon asymmetries are bounded by the positivity relations
%%%%%
\begin{equation}
|A_1| \leq 1 \; \; \; \; \; \; 
{\rm and} \; \; \; \; \; \; |A_2(x)| \leq \sqrt{R(x)} \; .
\label{eq:pos}
\end{equation}
%%%%%
Since also $\eta \ll 1$ in the kinematic range of most high energy experiments,
the term proportional to $A_2$ can be neglected, and
%%%%%
\begin{equation}
A_1 \simeq \frac{A_{LL}}{D} \; .
\label{eq:asym4}
\end{equation}
%%%%%

In {\tt POLDIS} both asymmetries, $A_{LL}$ and $A_1$, are generated.

%%%%%%%%%%
\subsection{
\label{sec:formulae}
Partonic cross sections}

Owing to factorization, the unpolarized (polarized) DIS cross section can be
written as a convolution of the unpolarized (polarized) parton distribution
function $F$ ($\Delta F$) with the partonic hard-scattering cross sections
${\rm d} {\hat \sigma}$ (${\rm d} \Delta {\hat \sigma}$):
%%%%%
\begin{equation}
{\rm d} \sigma^\lambda \sim F \otimes {\rm d} {\hat \sigma} \, + 
\lambda \, \Delta F \otimes {\rm d} \Delta {\hat \sigma} \; .
\label{eq:cs1}
\end{equation}
Here $\lambda$ refers to the parallel $\uparrow\uparrow$ ($\lambda = + 1$)
and antiparallel $\uparrow\downarrow$ ($\lambda = - 1$)
spin configuration of the incoming lepton and target nucleon in
the $\gamma^\ast - N$ c.m..

The terms ${\rm d} {\hat \sigma}$ and ${\rm d} \Delta {\hat \sigma}$ are the 
spin-independent 
%%%%%
\begin{equation}
{\rm d} {\hat \sigma} = \frac{1}{2} \,
({\rm d} {\hat \sigma}^{\uparrow\uparrow} +
{\rm d} {\hat \sigma}^{\uparrow\downarrow})
\label{cs3}
\end{equation}
%%%%%
and spin-dependent
%%%%%
\begin{equation}
{\rm d} \Delta {\hat \sigma} = \frac{1}{2} \, 
({\rm d} {\hat \sigma}^{\uparrow\uparrow} -
{\rm d} {\hat \sigma}^{\uparrow\downarrow}) \; .
\label{cs4}
\end{equation}
%%%%%
parts of the partonic hard cross section.
One introduces also the partonic asymmetry $\widehat{a}_{LL}$
for the hard-scattering process
%%%%%
\begin{equation}
\widehat{a}_{LL} = \frac{{\rm d} \Delta {\hat \sigma}}{{\rm d} {\hat \sigma}}
\; .
\label{cs5}
\end{equation}
%%%%%

The scattering asymmetry $A_{LL}$ for the reaction is obtained from
%%%%%
\begin{equation}
A_{LL} = \frac{\sum \int {\rm d} \Delta {\hat \sigma} \, \Delta F}
              {\sum \int {\rm d} {\hat \sigma} \, F}
\label{eq:asym5}
\end{equation}
%%%%%
where the sum runs over the hard-scattering sub-processes
calculated to first order in $\alpha_s$ 
(the corresponding Feynman diagrams for ${\rm d} {\hat \sigma}$ are shown
in Fig.~\ref{fig:feydia}),
and is integrated over the accessible phase space.
Using the partonic scattering asymmetry $\widehat{a}_{LL}$,
the asymmetry in Eq.~\ref{eq:asym5} can be rewritten, as
%%%%%
\begin{equation}
A_{LL} = \frac{\sum \int {\rm d} {\hat \sigma} F \,\, \widehat{a}_{LL} \,\, 
\Delta F / F}
{\sum \int {\rm d} {\hat \sigma} F}
\label{eq:asym6}
\end{equation}
%%%%%

In the Monte Carlo calculation,
${\rm d} {\hat \sigma} \, F$ can be viewed as the {\it cross section weight},
and $\widehat{a}_{LL} \,\Delta F / F$ as the {\it asymmetry weight}
for the generated event.
The integrals in Eq.~\ref{eq:asym6} are performed by summing these
quantities for all events in the generated sample.

%%%%%%%%%%
\subsection{
\label{sec:formulaeb}
Hard cross sections at order $\alpha_s$ in QCD}

%%%%%
\begin{figure}
\vspace*{-10mm}
\begin{center}
\mbox{\epsfxsize=16cm\epsffile{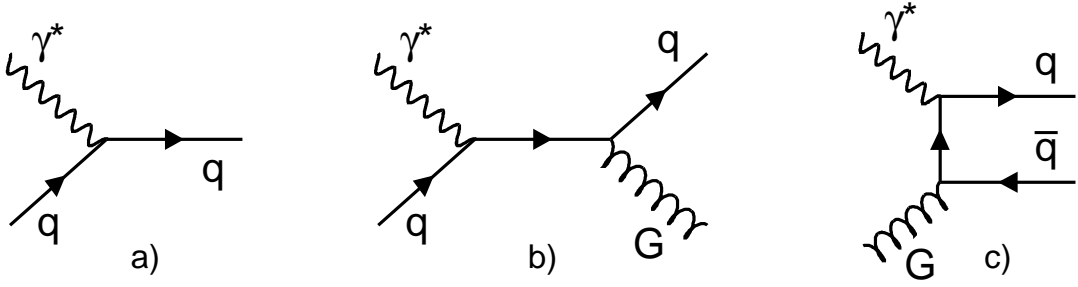}}
\end{center}
\vspace*{-5mm}
\caption{Lowest order Feynman diagrams for DIS: a) leading order, b) Compton,
c) Photon-gluon fusion.}
\label{fig:feydia}
\end{figure} 
%%%%%

The leading order (L.O.) parton level process is the virtual photo-absorption
$\gamma^\ast + q \rightarrow q$ (Fig.~\ref{fig:feydia}a).
At first order in QCD the gluon radiation (Compton diagram)
$\gamma^\ast + q \rightarrow q+G$ (Fig.~\ref{fig:feydia}b)
and the photon-gluon fusion (PGF) $\gamma^\ast + G \rightarrow q + \bar{q}$
(Fig.~\ref{fig:feydia}c) also contribute to the DIS cross section.
Since for the latter two processes there are two partons in the final state, 
the first order matrix elements involve three new degrees of freedom 
in addition to the two variables, say $x$ and $Q^2$,
needed for the L.O. DIS diagram (Fig.~\ref{fig:feydia}a).
These three new degrees of freedom correspond to the energy, polar angle,
and the azimuthal angle $\phi$ between the lepton and QCD scattering plane
of one of the final partons (the other is fixed by the kinematics).
A suitable choice for these new degrees of freedom is~\cite{Pec80}:
%%%%%
\begin{equation}
x_p = \frac{x}{\xi} \; , \; \; \; \; 
z_q = \frac{p \cdot p_q}{p \cdot q} \; , \; \; \; \;
\phi = \frac{(\vec{p} \times \vec{l}) \cdot (\vec{p} \times \vec{p_q})}
{|\vec{p} \times \vec{l}| |\vec{p} \times \vec{p_q}|}
\label{eq:var}
\end{equation}
%%%%%
where $\xi$ is the momentum fraction of the incoming parton, and
$p_q$ is the momentum of the final quark,
and the cross sections are five-fold differential
%%%%%
\begin{equation}
\frac{{\rm d}^5 {\hat \sigma} (x, Q^2, x_p, z_q, \phi)}
{{\rm d}x \, {\rm d}Q^2 \, {\rm d}x_p \, {\rm d}z_q \, {\rm d}\phi} \; .
\label{eq:fivefold}
\end{equation}
%%%%%

In the virtual boson-parton c.m.~frame the unpolarized 
cross section ${\rm d} {\hat \sigma}$ can be decomposed as~\cite{Pec80}:
%%%%%
\begin{equation}
{\rm d} {\hat \sigma} =
{\rm d} {\hat \sigma}_0 + \cos\phi \; {\rm d} {\hat \sigma}_1 + \cos2\phi \;
{\rm d} {\hat \sigma}_2
\label{eq:dec1}
\end{equation}
%%%%%
(note ${\rm d} {\hat \sigma}_i = {\rm d} {\hat \sigma}_i (x, Q^2, x_p, z_q)$)
and the polarized cross section ${\rm d} \Delta {\hat \sigma}$ as:
%%%%%
\begin{equation}
{\rm d} \Delta {\hat \sigma} =
{\rm d} \Delta {\hat \sigma}_0 + \cos\phi \; {\rm d} \Delta {\hat \sigma}_1 \; .
\label{eq:dec2}
\end{equation}
%%%%%
The $\cos 2 \phi$ term does not appear in Eq.~\ref{eq:dec2},
because it enters only in the cross section for the virtual photon
longitudinal component,
and therefore cancels in $\Delta {\hat \sigma}$. 
After integration over the azimuthal angle $\phi$, only the first
term on the right-hand side of Eqs.~\ref{eq:dec1} and~\ref{eq:dec2} remains
({\it i.e.} ${\rm d} {\hat \sigma}_0$ and ${\rm d} \Delta {\hat \sigma}_0$).

When studying HF production, the masses of the quarks must be taken
into account.
In this case the helicity does not coincide with the quark spin:
$q_{\pm1/2} = q_{R/L} + O \left( m / \sqrt{\hat{s}} \right)$
(the subscript $\pm 1/2$ denotes the quark helicity,
and $q_{R/L} = 1/2 (1 \pm \gamma_5) q$).
The HF photon-gluon fusion spin asymmetry $\widehat{a}_{LL}$
reaches the value $-1$ of the massless case
only in the asymptotic limit of very high energies,
while at threshold $\widehat{a}_{LL} = +1$.
In the HF lepto-production via the PGF the relevent scales of the process
are set by the HF quark mass $m_Q$, and, therefore,
this process can be studied also in the quasi-real photo-production limit
of $Q^2 \rightarrow 0$.

In Appendix~A we summarize these partonic cross sections calculated
for one photon exchange following the cross section decomposition of
Eqs.~\ref{eq:dec1} and~\ref{eq:dec2}.
In Appendix~B we summarize the same cross sections expressed in terms 
of the Mandelstam variables ${\hat s}$, ${\hat t}$, and ${\hat u}$
integrated over the azimuthal angle $\phi$.
The unpolarized cross sections have been derived in~\cite{Pec80} for the
massless case and in~\cite{Sch88} for HF. The polarized ones were
re-derived by us, extending also the results of~\cite{pepsi,Rat83,Wat82}.

\begin{figure}
\vspace*{-10mm}
\begin{center}
\mbox{\epsfxsize=16cm\epsffile{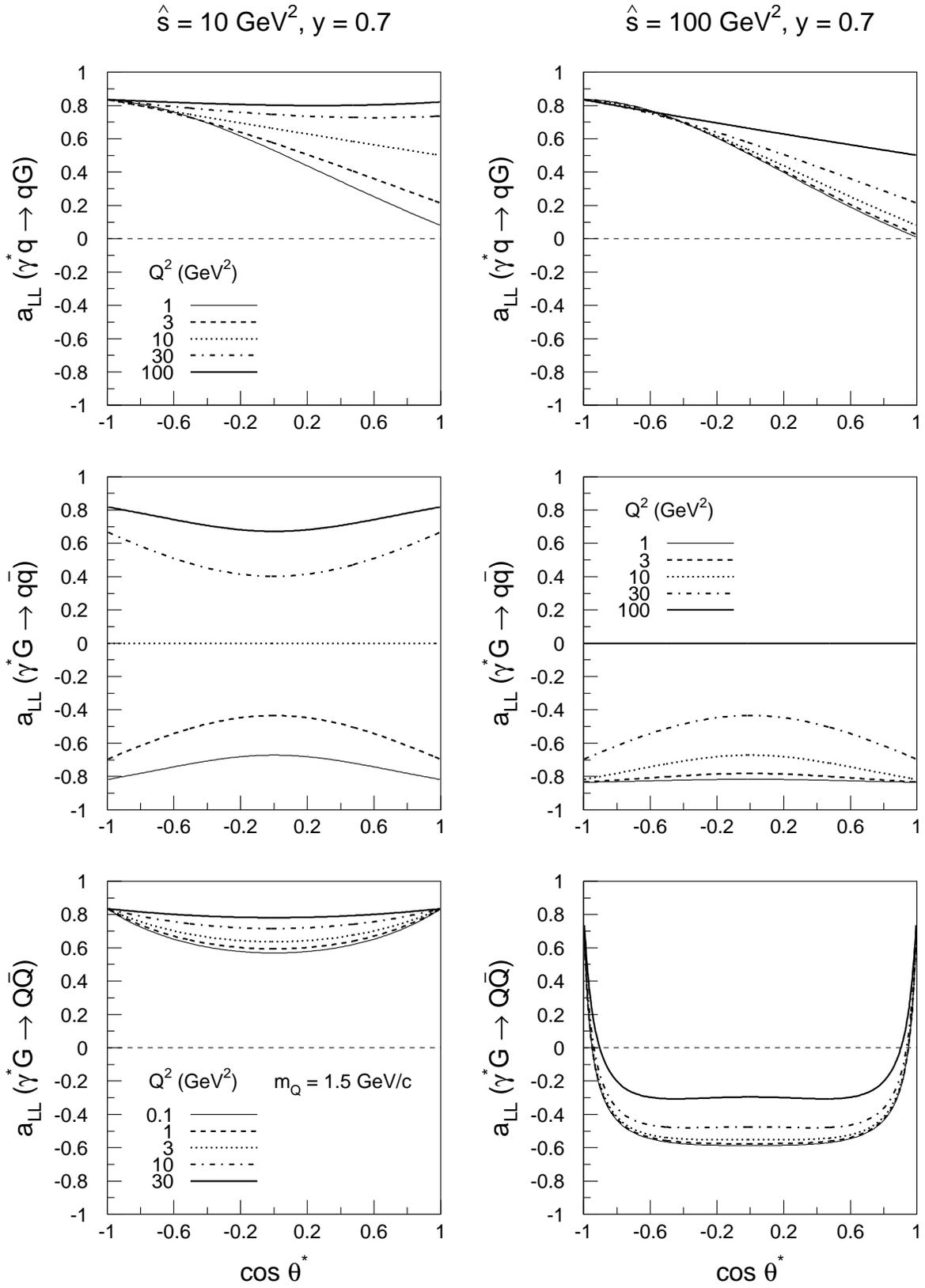}}
\end{center}
\vspace*{-5mm}
\caption{The scattering asymmetry $\widehat{a}_{LL}$ for
$\gamma^\ast + q \rightarrow q + G$,
$\gamma^\ast + G \rightarrow q + \bar{q}$, and
$\gamma^\ast + G \rightarrow Q + \bar{Q}$ (HF)
as a function of the c.m. scattering angle $\vartheta^\ast$ at fixed $y = 0.7$
for different values of $Q^2$ and for two values of the c.m. energy $\hat{s}$.}
\label{fig:asymm}
\end{figure}

The partonic scattering asymmetry $\widehat{a}_{LL}$ is obtained from
Eq.~\ref{cs5} by adding up the various terms of the cross sections 
in Eqs.~\ref{eq:dec1} and~\ref{eq:dec2}, which are summarized in Appendix~A.
For instance, the L.O. scattering asymmetry is
\begin{equation}
\widehat{a}_{LL}^{\gamma^\ast q \rightarrow q} \, = \, 
\frac{1-(1-y)^2}{1+(1-y)^2} \; .
\label{eq:all}
\end{equation}
For the other Feynman diagrams shown in Fig.~\ref{fig:feydia},
the partonic asymmetries result in much more complicated expressions.
In Figure~\ref{fig:asymm} we plot the partonic scattering asymmetries
$\widehat{a}_{LL}$ of order $\alpha_s$
as a function of the c.m. scattering angle 
$\vartheta^\ast$ between the incoming and outgoing partons for various values
of the $\gamma^\ast$ momentum transfer $Q^2$ and of the c.m. energy $\hat{s}$.
To be noted the $Q^2$ dependence of these asymmetries.
The angle $\vartheta^\ast$ is given by
%%%%%
\begin{equation}
\cos \vartheta^\ast = 1 - 2 z_q \; .
\label{eq:cos0}
\end{equation}
%%%%%

%%%%%
%%%%%

%%%%%%%%%%
\section{
\label{sec:MC}
Structure of the program}

{\tt POLDIS} consists of a set of subroutines
for the handling of the polarization,
which are linked with the unpolarized lepto-production event generetor 
{\tt LEPTO}~\cite{lepto} or {\tt AROMA}~\cite{aroma}.
The hadronization is performed with {\tt JETSET}~\cite{jetset}
using the LUND string model.
No modifications to these event generators is required.
The {\tt PDFLIB} library~\cite{pdflib} is used as a source for the
unpolarized parton distribution functions.
We assume familiarity with these programs.
Various polarized parton distribution functions, obtained from their authors,
are also included, although in a less standardized form.

The general structure of {\tt POLDIS} is similar to a typical Monte Carlo
program using {\tt LEPTO} or {\tt AROMA} with the addition 
of calls to some subroutines for the polarization calculations.
{\tt POLDIS} produces in output, in addition to the standard {\tt LEPTO} output,
the polarized scattering asymmetries $A_{LL}$ and $A_1$, and
the spin-dependent cross sections.
{\tt POLDIS} (as {\tt LEPTO}) is a {\it slave} program,
in the sense that the main {\it steering} code
for the administration of the event generation and
the subsequent analysis of these events,
has to be provided by the user.

The various relevant {\tt POLDIS} program components are summarized
in Tab.~\ref{tab:comp}. Most of names start with POL.
These are the only components that may be accessed by the user.
The program settings and parameters are listed in Tab.~\ref{tab:param}.
These parameters contains the values of different spin asymmetries,
which need to be accessed by the user.

%%%%%
\begin{table}
\begin{center}
\begin{tabular}{|l|p{12cm}|}
\hline
POLINI (S) & initializes the (un)polarized parton distribution functions \\
POLASYM (S) & calculates the {\it polarization weight} for each generated event
with calls to the functions for the calculation of the spin asymmetries
calls to the subroutines for the extraction of unpolarized
and polarized parton densities,
and the calculation of $R$ \\
POLSTR (S) & returns the values of the polarized parton densities at given
$x$ and $Q^2$ (PDG flavor code convention) \\
POLINTL (S) & contains the internal set of polarized parton densities \\
POLEND (S) & gives the spin-dependent cross sections (Monte Carlo estimate) \\
ALLQ (F) & calculates $\widehat{a}_{LL}$ for $\gamma + q \rightarrow q$ \\
ALLQG (F) & calculates $\widehat{a}_{LL}$ for $\gamma + q \rightarrow q + G$ \\
ALLQQ (F) & calculates $\widehat{a}_{LL}$ for 
$\gamma + G \rightarrow q + \bar{q}$ \\
ALLQQHF (F) & calculates $\widehat{a}_{LL}$ for
$\gamma + G \rightarrow Q + \bar{Q}$ \\
RPAR (F) & gives the value of $R$ at given $x$ and $Q^2$ \\
POLDISU (C) & contains the {\tt POLDIS} settings and parameters which include
also the asymmetry values \\
\hline
\end{tabular}
\end{center}
\caption{Relevant {\tt POLDIS} program components: subroutines~(S),
functions~(F), and common blocks~(C).}
\label{tab:comp}
\end{table}
%%%%%

\begin{itemize}
\item 
At the {\bf initialization stage} standard
{\tt LEPTO} and/or {\tt AROMA} parameters and switches are selected.
Additionally, a parametrization for the polarized distribution functions
and for $R = \sigma^L  / \sigma^T $ are chosen,
and a kinematical interval is defined for the {\it simulation}.
In the subroutine {\bf polini} the selected polarized distribution
functions are read from the corresponding ASCII file(s).
Immediately after that the unpolarized event generator is initialized
with a call to {\bf linit} ({\tt LEPTO}) or {\bf arinit} ({\tt AROMA}).
\item
In the {\bf event loop} unpolarized events are generated with calls to
{\bf lepto} (or {\bf aroma}).
The asymmetry is calculated for each event in the subroutine {\bf polasym},
and the asymmetry results are stored in the common block {\bf /poldisu/}.
At this point the user can perform additional analysis on the generated event:
for instance, he can select a binning {\it e.g.} in $x$ or $y$
for the asymmetries $A_{LL}$ and $A_1$,
or study semi-inclusive asymmetries by requiring in the event a $\pi^+$
with $z > 0.2$.
\item
In the {\bf ending stage} the spin-dependent cross sections
are estimated in the subroutine {\bf polend}.
The relevant results are printed in the form of a table.
\end{itemize}

%%%%%
\begin{table}
\begin{center}
\begin{tabular}{|l|l|}
\hline
POLLST(1) & polarized parton distribution function \\
POLLST(2,3) &
unpolarized parton distribution function (in {\tt PDFLIB} format) \\
POLLST(4) & parametrization of $R$ \\
POLLST(5-9) & unused at present \\
POLLST(10) & number of generated events used in the asymmetry calculation \\
\hline
POLPAR(1) & $\widehat{a}_{LL}$ for current event \\
POLPAR(2) & $R$ ($\sigma^L / \sigma^T$) for current event \\
POLPAR(3) & $D$ (depolarization) for current event \\
POLPAR(4-6) & $\Delta F / F$ for current event for the subsets \\
POLPAR(11-13) & $A_{LL}$ for current event \\
POLPAR(14-16) & $A_1$ for current event \\
POLPAR(17-19) & $A_{LL}$ for the generated sample \\
POLPAR(20-22) & $A_1$ for the generated sample \\
POLPAR(23-28) & unused at present \\
POLPAR(29-31) &
$\sigma^{\uparrow\uparrow}$ in pb --
Monte Carlo estimate associated with generated event sample \\
POLPAR(32-34) &
$\sigma^{\uparrow\downarrow}$ in pb \\
POLPAR(35-40) & unused at present \\
\hline
\end{tabular}
\end{center}
\caption{{\tt POLDIS} parameters in common block {\bf /poldisu/}:
POLLST is an array of integers,
and POLPAR an array of double precision real numbers.}
\label{tab:param}
\end{table}
%%%%%

%%%%%%%%%%
\subsection{
\label{sec:asymmcalc}
Asymmetry evaluation}

The scattering asymmetry $\widehat{a}_{LL}$ is calculated
for each generated event according to the underlying sub-process
and the kinematic variables given by the unpolarized event generator.
The unpolarized ($F$) parton densities are evaluated at the given 
$x$ and $Q^2$.
The polarized ($\Delta F$) parton densities are also evaluated at the given 
$x$ and $Q^2$ for 2 or 3 subsets of the selected polarized 
parton distribution functions set
(see Section~\ref{sec:polpdf}),
and the parton polarizations, $\Delta F / F$, are thus obtained.
The {\it polarization weight}
\begin{equation}
\widehat{w}_{LL} = \widehat{a}_{LL} \times \frac{\Delta F}{F}
\end{equation}
%%%%%
is finally calulated for all subsets.

The values of the asymmetries $A_{LL}$ and $A_1$ are updated event by event:
%%%%%
\begin{equation}
A_{LL} = \frac{N-1}{N} A_{LL} + \frac{1}{N} \, \widehat{w}_{LL} \; \; \; \; \;
{\rm and} \; \; \; \; \;
A_1 = \frac{N-1}{N} A_1 + \frac{1}{N} \frac{\widehat{w}_{LL}}{D} \; .
\end{equation}
%%%%%
$N$ is the number of events generated so far, and
$D$ is the virtual photon depolarization (Eq.~\ref{eq:depol}).
The values of the asymmetries corresponding to different $\Delta F$'s
from the same subset are stored in the common block {\bf /poldisu/}.
The statistical accuracy on the asymmetries depends on the number of
generated events $N$ and goes as $1 / \sqrt{N}$.
To study, for instance, the $x$ and/or $Q^2$ behavior of the asymmetry,
the values of $A_{LL}$ and $A_1$ can be binned as a function of
$x$ and/or $Q^2$.

The spin-dependent cross sections are obtained from
the unpolarized cross section $\sigma^0$
and the scattering asymmetry $A_{LL}$:
\begin{eqnarray}
\sigma^{\uparrow \uparrow} & = & \frac{1}{2} \, \sigma^0 \, (1 + A_{LL}) \\
\sigma^{\uparrow \downarrow} & = & \frac{1}{2} \, \sigma^0 \, (1 - A_{LL})
\end{eqnarray}
 
%%%%%%%%%%
\subsection{
\label{sec:polpdf}
Polarized parton distribution functions}

The following polarized parton distribution functions for the proton
are presently included
(in parenthesis are shown the corresponding unpolarized parton densities,
which should be used for consistency and are automatically selected,
and the $x / Q^2$ range of validity of the parametrization;
the notation adopted is that of {\tt PDFLIB}):
\begin{itemize}
\item[1 ] GS-95 (unpolarized: DO 1.1; 
range: $x > 10^{-5}$, $4 < Q^2 < 4.5 \times 10^5~{\rm GeV}^2$)~\cite{GS95}
\item[2 ] GS-96LO (unpolarized: GRV-94LO;
range: $x > 10^{-5}$, $1 < Q^2 < 10^6~{\rm GeV}^2$)~\cite{GS96}
\item[3 ] GS-96NLO (unpolarized: MRSA$^\prime$;
range: $x > 10^{-5}$, $1 < Q^2 < 10^6~{\rm GeV}^2$)~\cite{GS96} 
\item[4 ] GRSV-96LO (unpolarized GRV-94LO;
range: $x > 10^{-4}$, $0.4 < Q^2 < 10^4~{\rm GeV}^2$)~\cite{GRSV}
\item[5 ] GRSV-96NLO (unpolarized GRV-94HOMS;
range: $x > 10^{-4}$, $0.4 < Q^2 < 10^4~{\rm GeV}^2$)~\cite{GRSV}
\end{itemize}
Typically, each set of polarized parton densities contain two or three 
different parametrizations, obtained from the same analysis.
The GS polarized parton distribution functions contain three
subsets, referred as {\it set~A}, {\it set~B}, and {\it set~C}.
The GRSV polarized parton distribution functions contain two
subsets, referred as {\it standard} and {\it valence scenario}.
In {\tt POLDIS} all subsets are used simultaneously,
thus giving an output with two or three values for the spin asymmetries
$A_{LL}$ and $A_1$ corresponding to the used subsets.
The parton distribution functions are stored on a $x / Q^2$ grid
for each parton component and stored in ASCII files.
These polarized parton distribution functions have been obtained from the 
corresponding authors~\cite{GS95,GS96,GRSV}.

An internal set of polarized distribution functions is also included,
mainly for debugging and apparatus studies.
Different parametrizations can be implemented by the user by simply editing
the subroutine {\bf polintl}, which contains this internal set.
This internal set can be also used, for instance,
for evaluating the effects of a large
negative sea polarization $\Delta {\sf s} < 0$ on the scattering asymmetry
for a particular channel.
It is assumed that these polarized parton densities scale as the
corresponding unpolarized distribution functions
(no dynamical generation of the sea and gluon polarization is performed).
For instance, the following parametrization, based on the
SU(6) spin structure of the proton combined with a {\it soft} gluon
can be used (and it is included):

%\begin{enumerate}
%\item[a ] 
$\Delta {\sf u_v} = {\sf u _v} - \frac{2}{3}{\sf d _v}, \,
\Delta {\sf d _v} = - \frac{1}{3}{\sf d _v}, \,
\Delta {\sf q_s} = 0, \, \Delta G = x G $
%\end{enumerate}

For {\it complex} targets ({\it i.e.} containing several protons
and neutrons) full isospin symmetry is assumed between protons and neutrons:
$\Delta {\sf u_v^p} = \Delta {\sf d_v^n}$,
$\Delta G^{\sf p} = \Delta G^{\sf n}$, etc., and possible nuclear effects
are neglected.

The selection of a polarized parton distribution function set is performed
by setting the corresponding switch to the desired value:
1 to 5 for the polarized parton densities listed above,
0 for the internal set, and $-1$ for no polarization.
At the initialization stage in subroutine {\bf polini},
in addition to the polarized parton density,
the corresponding unpolarized one is also selected.
Our default polarized parton distribution functions set is the GS-96LO,
combined with the unpolarized ones of GRV-94LO.

%%%%%%%%%%
\subsection{
\label{sec:rpar}
Parametrizations of R}

In most polarized DIS experiments the virtual photon asymmetry $A_1$ 
is obtained from the measured asymmetry $A_{LL}$ (Eq.~\ref{eq:asym4},
using a parametrization of $R = \sigma^L / \sigma^T$ determined
from unpolarized DIS data.
A similar approach is also adopted in {\tt POLDIS}.
The following parametrizations are included
(in parenthesis is shown the kinematic range over which $R$ was estimated,
and corresponds also to the region,
where the parametrization can be used safely):
\begin{itemize}
\item[1 ] SLAC (range: $x > 0.03$, $Q^2 > 0.35~{\rm GeV}^2$)~\cite{slac}
\item[2 ] NMC-97 (range: $x > 0.003$, $Q^2 > 0.30~{\rm GeV}^2$)~\cite{nmc97}
\item[3 ] BKS-97
(range: $4 \times 10^{-5} < x < 0.1$, 
$0.01 < Q^2 < 360~{\rm GeV}^2$)~\cite{bks97} 
\end{itemize}

The selection of the desired parametrization of $R$
is obtained by setting the corresponding switch to the appropriate value:
1 to 3 for the parametrizations listed above, and $0$ for $R = 0$.
Our default is the NMC-97 parametrization.

%%%%%%%%%%
\subsection{
\label{sec:comp}
{\tt POLDIS} parameters}

The common block {\bf /poldisu/} contains the program settings and
parameters as illustrtated in Tab.~\ref{tab:param}.

\begin{center}
COMMON / {\bf POLDISU} / POLLST(10), POLPAR(40)
\end{center}

In this common block different asymmetry values are stored.
For each spin asymmetry and spin-dependent
cross section there are three different values, which correspond to the
three subsets of the selected polarized parton distribution functions set.

%%%%%%%%%%
\subsection{
\label{sec:leptoint}
Interface with {\tt LEPTO}}

The relevant kinematics, program parameters, settings, and switches,
used also for the asymmetry calculations,
are stored in the {\tt LEPTO} common block {\bf /leptou/}:
\begin{center}
COMMON / {\bf LEPTOU} / CUT(14), LST(40), PARL(30), X, Y, W2, Q2, U
\end{center}
The correspondence between the kinematical variables stored in
the common block {\bf /leptou/} to the ones discussed in the previous
pages is:
\begin{center}
${\rm X} \equiv x$, ${\rm Q2} \equiv Q^2$, ${\rm Y} \equiv y$,
${\rm U} \equiv \nu$, and \\
PARL(28) = $x_p$, PARL(29) = $z_q$, PARL(30) = $\phi$.
\end{center}
Additionally, the following {\tt LEPTO} switches are used
for the asymmetry evaluation:

\noindent LST(22) specifies the struck nucleon: 
$1=$ proton, $2=$ neutron.

\noindent LST(24) specifies the hard-scattering sub-process:
$1=q$, $2=qG$, $3=q\bar{q}$, $5=Q\bar{Q}-$event (HF).

\noindent LST(25) specifies the struck quark:
$1={\sf d}$, $2={\sf u}$, $3={\sf s}$,
$-1={\sf \bar{d}}$, $-2={\sf \bar{u}}$, $-3={\sf \bar{s}}$.

The unpolarized cross sections measured in pb are stored in
PARL(23) (numerical integration at the initialization stage)
and in PARL(24) (Monte Carlo estimate).

The kinematics of the interaction and of all produced particles is stored
in the {\tt JETSET} common block {\bf /lujets/}.

%%%%%%%%%%
\section{
\label{sec:program}
How to run {\tt POLDIS}}

%%%%%
\begin{figure}
\begin{center}
\mbox{\epsfxsize=17cm\epsffile{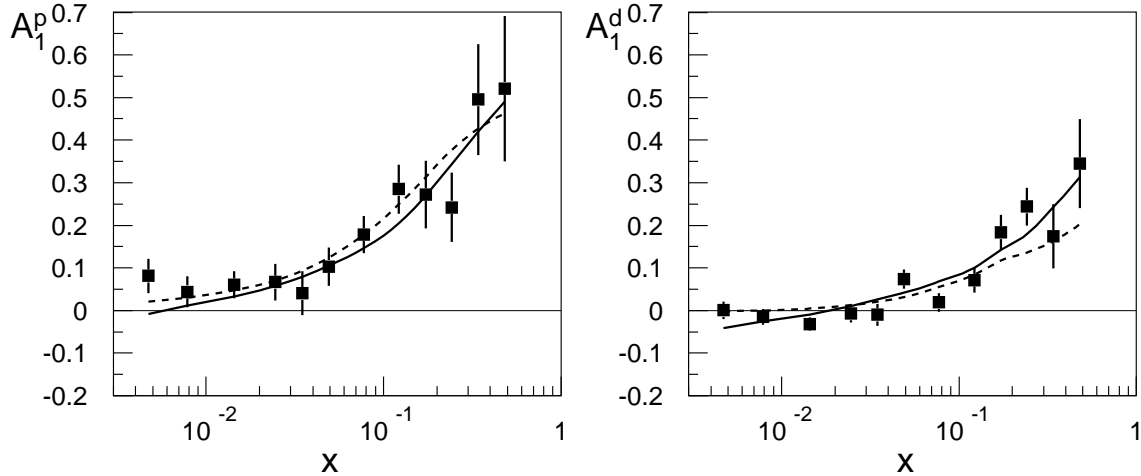}}
\end{center}
\vspace*{-10mm}
\caption{a) {\it Simulated} inclusive asymmetry $A_1^{\sf p}$
compared to SMC data from polarized protons~\protect\cite{smcp93}
using the GS-96LO {\it set~A} (full line) and the GRVS-96LO {\it standard scen.} (dashed line) polarized parton densities and the NMC-97 parametrization of $R$.
b) Same as (a) for $A_1^{\sf d}$ compared to SMC data
from a polarized deuteron target~\protect\cite{smcd95}.}
\label{fig:inc}
\end{figure} 
%%%%%

In addition to the standard {\tt LEPTO} (and {\tt AROMA}) input 
parameters and switches, such as the beam energy, target material, etc. 
(we assume the user to be familiar with them),
two additional input switches are required (see previous Section):

\noindent POLLST(1) = 0 to 5 for the polarized parton densities, and

\noindent POLLST(4) = 0 to 3 for the parametrization of $R$.

As already mentioned above, the user must provide a {\it steering} code
for the administration of the event generation and analysis.
Before the initialization of {\tt POLDIS} and {\tt LEPTO} the relevant
parameters, switches, etc. must be set to the corresponding values.

At the end of the event generation loop POLPAR contains various asymmetry
values and the spin-dependent cross sections associated with the subsets
of the selected polarized parton density set.

%%%%%%%%%%
\section{
\label{sec:test}
Results of test runs}

%%%%%
\begin{figure}
\begin{center}
\mbox{\epsfxsize=17cm\epsffile{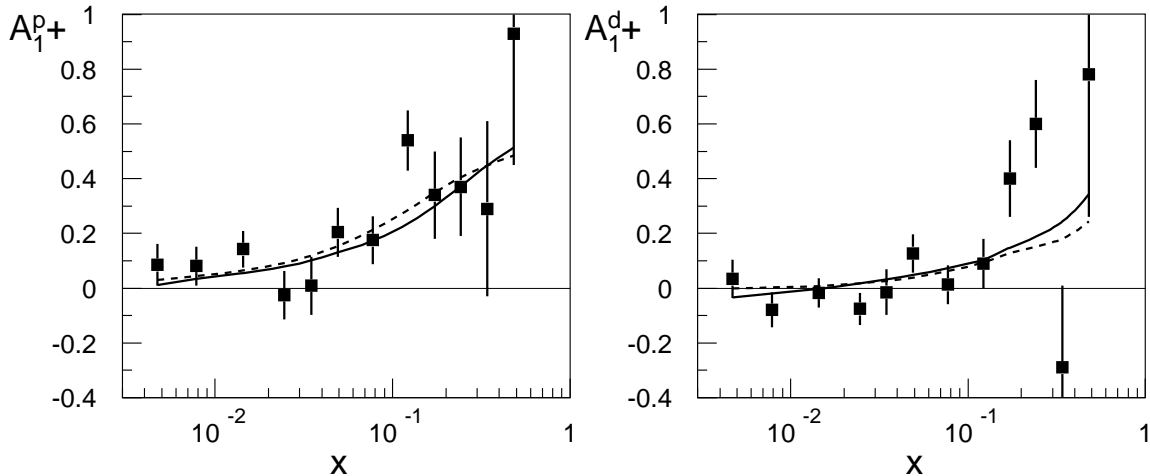}}
\end{center}
\vspace*{-10mm}
\caption{{\it Simulated} semi-inclusive asymmetry $A_{1, +}^{\sf p}$
on polarized protons (a) and $A_{1, +}^{\sf d}$ on polarized deuterons (b)
for positive hadrons with $z > 0.2$,
compared to SMC experimental data~\protect\cite{Ade96}
(GS-96LO {\it set~A} full line,
and GRVS-96LO {\it standard scen.} dashed line).}
\label{fig:sincp}
\end{figure} 
%%%%%

The inclusive asymmetry $A_1$, obtained with the
GS-96LO {\it set~A} and the GRVS-96LO {\it standard scen.}
polarized parton densities and the NMC-97 parametrization of $R$,
is compared in Fig.~\ref{fig:inc} to the SMC data
from polarized proton~\cite{smcp93} and deuteron~\cite{smcd95} targets.
A fairly accurate agreement between the {\it simulated} and the real data
can be observed in these plots.
It has to be noted, however, that polarized DIS data were used for the
evaluation of the polarized parton distribution functions used here,
and therefore this agreement can not be considered as a meaningful
physics result.
On the other hand, such an agreement shows the validity of the procedure
adopted in the simulation and the correctness of the calculations. 

\begin{figure}
\begin{center}
\mbox{\epsfxsize=17cm\epsffile{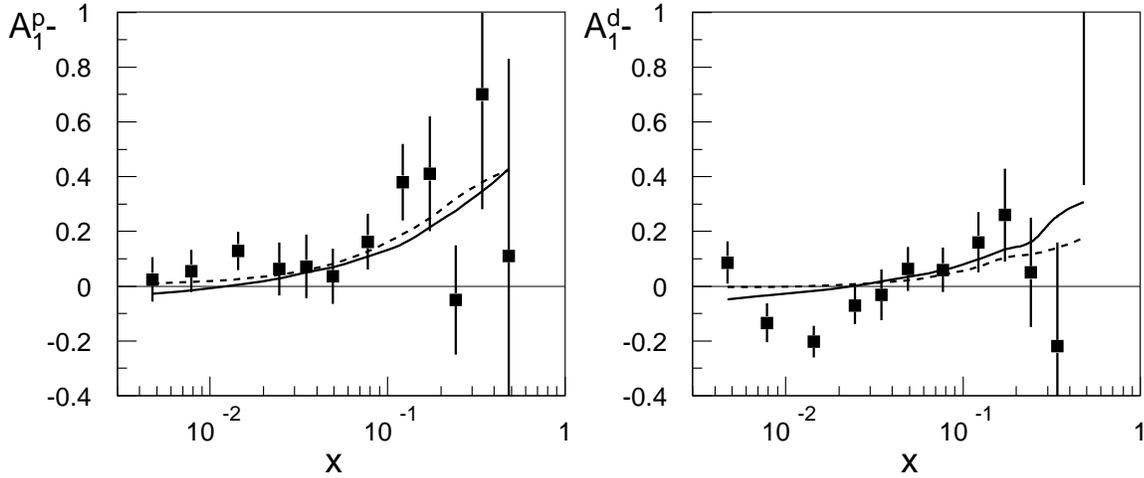}}
\end{center}
\vspace*{-10mm}
\caption{{\it Simulated} semi-inclusive asymmetry $A_{1, -}^{\sf p}$
on polarized protons (a) and $A_{1, -}^{\sf d}$ on polarized deuterons (b)
for negative hadrons with $z > 0.2$
(GRVS-96LO {\it standard} AND NMC-97),
compared to SMC experimental data~\protect\cite{Ade96}
(GS-96LO {\it set~A} full line,
and GRVS-96LO {\it standard scen.} dashed line).}
\label{fig:sincn}
\end{figure}

Figures~\ref{fig:sincp} and~\ref{fig:sincn} show the semi-inclusive
asymmetries for positive and negative hadrons generated with {\tt POLDIS}, respectively,
compared to the SMC data~\cite{Ade96} from polarized protons and deuterons.
Also for this {\it simulation} we used the GS-96LO {\it set~A} and
the GRVS-96LO {\it standard scen.}
polarized parton densities and the NMC-97 parametrization of $R$.
In the Monte Carlo simulation a separation between charged pions and kaons
can also be made.

%%%%%
%%%%%

%%%%%%%%%%
\section*{
\label{sec:ack}
Acknowledgments}

We would like to acknowledge G. Ingelman for discussions
on the LUND event generators used for this work,
and G.K.~Mallot for useful discussions on DIS.
We would like to thank T.~Gehrmann and V.~Vogelsang for providing us with their
polarized parton distribution functions (GS-95LO, GS-96LO, GS-96NLO, and
GRSV-96LO, GRSV-96NLO, respectively),
and B.~Badelek for providing us with the BKS-97 parametrization of $R$.
We would like also to thank E.~Rondio for using the preliminary versions
of this program.
This work is partially supported by KBN SPUB/P03/114/96.

%%%%%%%%%%
%\newpage

%%%%%%%%%%
\setcounter{equation}{0}
\renewcommand{\theequation}{A.\arabic{equation}}
\section*{
\label{sec:appa}
Appendix A: spin-averaged and spin-dependent hard cross sections}

In this Appendix we summarize the various terms appearing in the
unpolarized and polarized partonic level cross sections
(Eqs.~\ref{eq:dec1} and~\ref{eq:dec2})
for the processes included in this program.
To shorten the notation, the common factor 
\begin{equation}
\frac{2\alpha^2 e_q^2}{x Q^2}
\label{eq:emcpl}
\end{equation} 
describing the electromagnetic coupling is singled out.
$e_q$ is the charge of the struck quark.

\vspace*{6mm}
\noindent $\bullet \:\: \gamma^\ast \, q \rightarrow q$:

For the L.O. diagram (Fig.~\ref{fig:feydia}a),
the unpolarized cross section is:
%%%%%
\begin{eqnarray}
{\rm d} {\hat \sigma}_0 & = & \frac{1}{2} \, (1+(1-y)^2) \; ,
\end{eqnarray}
%%%%%
and the polarized one is:
%%%%%
\begin{eqnarray}
{\rm d} \Delta {\hat \sigma}_0 & = & \frac{1}{2} \, (1-(1-y)^2) \; .
\end{eqnarray}
%%%%%
Note that there is no dependence on the azimuthal angle $\phi$.

\vspace*{6mm}
\noindent $\bullet \:\: \gamma^\ast \, q \rightarrow q+g$:

For the Compton diagram  (Fig.~\ref{fig:feydia}b)
the unpolarized cross section is
%%%%%
\begin{eqnarray}
{\rm d} {\hat \sigma}_0 & = & \frac{2\alpha_s}{3\pi} \,
\left\{ \frac{1+(1-y)^2}{2} \, \left[ \frac{x_p^2+z_q^2}{(1-x_p)(1-z_q)}+2 \, 
(x_p z_q +1) \right] + (1-y) \, 4 x_p z_q \right\} \\
{\rm d} {\hat \sigma}_1 & = & \frac{4\alpha_s}{3\pi} \,
(y-2) \, \sqrt{1-y} \, \sqrt{\frac{x_p z_q}{(1-x_p)(1-z_q)}} \, 
(1 - x_p - z_q + 2 x_p z_q) \\
{\rm d} {\hat \sigma}_2 & = & \frac{4\alpha_s}{3\pi} \, (1-y) \, x_p z_q \; ,
\end{eqnarray}
%%%%%
and the polarized one is:
%%%%%
\begin{eqnarray}
{\rm d} \Delta {\hat \sigma}_0 & = & \frac{2\alpha_s}{3\pi} \,
\frac{1-(1-y)^2}{2} \,
\left[ \frac{x_p^2+z_q^2}{(1-x_p)(1-z_q)}+2 \, (x_p+z_q) \right] \\
{\rm d} \Delta {\hat \sigma}_1 & = & \frac{4\alpha_s}{3\pi} \,
y \, \sqrt{1-y} \, \sqrt{\frac{x_p z_q}{(1-x_p)(1-z_q)}} \, (1-x_p-z_q) \; .
\end{eqnarray}
%%%%%

\vspace*{6mm}
\noindent $\bullet \:\: \gamma^\ast \,g \rightarrow q+\bar{q}$
(light quarks):

For the PGF diagram (Fig.~\ref{fig:feydia}c) -- 
massless case ($m_q = 0$, \qq~= \uq, \dq, \sq)
the unpolarized cross section is:
%%%%%
\begin{eqnarray}
{\rm d} {\hat \sigma}_0 & = & \frac{\alpha_s}{4\pi} \, 
\left\{ \frac{1+(1-y)^2}{2} \, \frac{\left[ x_p^2+(1-x_p)^2 \right] 
\left[ z_q^2+(1-z_q)^2 \right]}
{z_q (1-z_q)} + (1-y) \, 8 x_p (1-x_p) \right\} \\
{\rm d} {\hat \sigma}_1 & = & \frac{\alpha_s}{2\pi} \,
(y-2) \, \sqrt{1-y} \, \sqrt{\frac{x_p(1-x_p)}{z_q(1-z_q)}} \, 
(1-2x_p) \, (1-2z_q) \\
{\rm d} {\hat \sigma}_2 & = & \frac{\alpha_s}{\pi} \,
(1-y) \, x_p (1-x_p) \; ,
\end{eqnarray}
%%%%%
and the polarized one is:
%%%%%
\begin{eqnarray}
{\rm d} \Delta {\hat \sigma}_0 & = & \frac{\alpha_s}{4\pi} \, \frac{1-(1-y)^2}{2} \,
\frac{(2 x_p-1) \left[ z_q^2+(1-z_q)^2 \right] }
{z_q (1-z_q)} \\ 
{\rm d} \Delta {\hat \sigma}_1 & = & \frac{\alpha_s}{2\pi} \, 
y \, \sqrt{1-y} \, \sqrt{\frac{x_p(1-x_p)}{z_q(1-z_q)}} \, (1-2z_q) \; . 
\end{eqnarray}
%%%%%

\vspace*{6mm}
\noindent $\bullet \:\: \gamma^\ast \, g \rightarrow Q+\bar{Q}$
(heavy quarks):

For the PGF diagram -- 
massive case ($m_Q \neq 0$, \QQ~= \cq, \bq)
the unpolarized cross section is:
\begin{eqnarray}
{\rm d} {\hat \sigma}_0 & = & \frac{\alpha_s}{4\pi} \,
\left\{ \frac{1+(1-y)^2}{2} \, \left[ \frac{\left[ x_p^2+(1-x_p)^2 \right]
\left[ z_q^2+(1-z_q)^2 \right] + \beta (1 - 2 x_p)}{z_q (1-z_q)} + 
\frac{\beta (2 x_p - \beta) }{4 z_q^2 (1-z_q)^2}\right] \right. \nonumber \\
& & \left. + \, (1-y) \,
\left[ 8 x_p (1-x_p) - \frac{2 \beta x_p }{z_q (1-z_q)}\right]\right\} \\
{\rm d} {\hat \sigma}_1 & = & \frac{\alpha_s}{2\pi} \, (y -2) \, \sqrt{1-y} \, 
\sqrt{\frac{x_p (1-x_p)}{z_q(1-z_q)} - \frac{\beta x_p } {4 z_q^2 (1-z_q)^2}}
\nonumber \\
& & \cdot \, (1-2z_q) \left( 1-2x_p - \frac{\beta}{2 z_q (1-z_q)} \right) \\   
{\rm d} {\hat \sigma}_2 & = & \frac{\alpha_s}{\pi} \,
(1-y) \, \left[ x_p (1-x_p) + \frac{\beta (1 - 2 x_p)}{4 z_q (1-z_q)} - 
\frac{\beta^2}{16 z_q^2 (1-z_q)^2}\right] \; ,
\label{mfus1}
\end{eqnarray}
%%%%%
and the polarized one is:
%%%%%
\begin{eqnarray}
{\rm d} \Delta {\hat \sigma}_0 & = & \frac{\alpha_s}{4\pi} \,
\frac{1-(1-y)^2}{2} 
\left[ \frac{2 x_p-1}{z_q (1-z_q)} + \frac{\beta}{2 z_q^2 (1-z_q)^2} \right] 
\left[ z_q^2+(1-z_q)^2 \right] \\
{\rm d} \Delta {\hat \sigma}_1 & = & \frac{\alpha_s}{2\pi} \, y \,
\sqrt{1-y} \, \sqrt{\frac{x_p (1-x_p)}{z_q (1-z_q)}-
\frac{\beta x_p }{4 z_q^2} (1-z_q)^2} \, (1-2 z_q) \; .
\label{mfus2}
\end{eqnarray}
The factor $\beta$ is related to the HF quark velocity $v$ in the
photon-gluon c.m. by $v = \sqrt{1-\beta}$, and it is given by
%%%%%
\begin{equation}
\beta = 4 x_p \frac{m^2_Q}{Q^2}
\label{beta}
\end{equation}
%%%%%
where $m_Q$ is the mass of the heavy quark.

%%%%%%%%%%
\setcounter{equation}{0}
\renewcommand{\theequation}{B.\arabic{equation}}
\section*{
\label{sec:appb}
Appendix B: spin-averaged and spin-dependent hard cross sections
using Mandelstam variables}

In this Appendix we rewrite the unpolarized and polarized partonic
cross sections (Eqs.~\ref{eq:dec1} and~\ref{eq:dec2})
integrated over the azimuthal angle $\phi$
in terms of the Mandelstam variables $\hat{s}$, $\hat{t}$, and $\hat{u}$.
The relations between these variables and the $x_p$, $z_q$ variables 
used in Appendix~A are:
%%%%%
\begin{eqnarray}
\hat{s} & = & Q^2 \frac{1 - x_p}{x_p} \; = \; 2M\nu\xi - Q^2 \nonumber \\
\hat{u} & = & m_Q^2 -Q^2 \frac{z_q}{x_p} \\
\hat{t} & = & m_Q^2 -Q^2 \frac{1-z_q}{x_p} \nonumber \\
\hat{s} \, + \, \hat{t} \, + \, \hat{u} & = & 2m_Q^2 -Q^2 \nonumber
\end{eqnarray}
%%%%%
where $\xi$ is the momentum fraction of the incident parton 
(as in Eq.~\ref{eq:var}) and $m_Q$ is the mass of the (HF) quark.

These cross sections expressed in terms of the Mandelstam variables
depend on $Q^2$, $x$, $\hat{s}$, and $\hat{t}$
(${\rm d} \hat{\sigma}_i = {\rm d} \hat{\sigma}_i (x, Q^2, \hat{s}, \hat{t}$).
The common factor describing the electromagnetic coupling
%%%%%
\begin{equation}
\frac{4 \pi \alpha^2 e_q^2}{x Q^2}
\end{equation} 
%%%%%
is singled out as in Appendix~A. The additional factor $2\pi$ compared to
Eq.~\ref{eq:emcpl} comes from the integration in $\phi$.

\vspace*{6mm}
\noindent $\bullet \:\: \gamma^\ast \, q \rightarrow q$:

For the L.O. diagram (Fig.~\ref{fig:feydia}a),
the unpolarized cross section is:
%%%%%
\begin{eqnarray}
{\rm d} {\hat \sigma}_0 & = & \frac{1}{2} \, (1+(1-y)^2) \; ,
\end{eqnarray}
%%%%%
and the polarized one is:
%%%%%
\begin{eqnarray}
{\rm d} \Delta {\hat \sigma}_0 & = & \frac{1}{2} \, (1-(1-y)^2) \; .
\end{eqnarray}
%%%%%

\vspace*{6mm}
\noindent $\bullet \:\: \gamma^\ast \, q \rightarrow q+g$:

For the Compton diagram  (Fig.~\ref{fig:feydia}b)
the unpolarized cross section is:
%%%%%
\begin{eqnarray}
{\rm d} {\hat \sigma}_0 & = &
\frac{2\alpha_s}{3\pi} \frac{1}{(\hat{s} + Q^2)^2} \nonumber \\
& & \cdot \left\{ \frac{1+(1-y)^2}{2} \,
\left[ 2 - \frac{2 \, \hat{u} \, Q^2}{(\hat{s} + Q^2)^2} -
\frac{Q^4 + \hat{u}^2}{\hat{s} \, \hat{t}}\right] -
(1-y) \, \frac{4 \, Q^2 \hat{u}}{(\hat{s} + Q^2)^2}\right\} \; ,
\end{eqnarray}
%%%%%
and the polarized one is:
%%%%%
\begin{eqnarray}
{\rm d} \Delta {\hat \sigma}_0 & = &
\frac{2\alpha_s}{3\pi} \frac{1}{(\hat{s} + Q^2)^2} \,
\frac{1-(1-y)^2}{2} \,
\left[ \frac{2 \, (Q^2-\hat{u})}{\hat{s}+Q^2} - 
\frac{Q^4+\hat{u}^2}{\hat{s} \, \hat{t}} \right]\; .
\end{eqnarray}

\vspace*{6mm}
\noindent $\bullet \:\: \gamma^\ast \,g \rightarrow q+\bar{q}$
(light quarks):

For the PGF diagram (Fig.~\ref{fig:feydia}c) -- 
massless case ($m_q = 0$, \qq~= \uq, \dq, \sq)
the unpolarized cross section is:
%%%%%
\begin{eqnarray}
{\rm d} {\hat \sigma}_0 & = &
\frac{\alpha_s}{4\pi} \frac{1}{(\hat{s} + Q^2)^2} \, 
\left\{ \frac{1+(1-y)^2}{2} \, \frac{Q^4 + \hat{s}^2}{(\hat{s} + Q^2)^2} \,
\frac{\hat{u}^2 + \hat{t}^2}{\hat{u} \, \hat{t}} + (1-y) \, 
\frac{8 \, Q^2 \hat{s}}{(\hat{s} + Q^2)^2} \right\} \; ,
\end{eqnarray}
%%%%%
and the polarized one is:
%%%%%
\begin{eqnarray}
{\rm d} \Delta {\hat \sigma}_0 & = &
\frac{\alpha_s}{4\pi} \frac{1}{(\hat{s} + Q^2)^2} \,
\frac{1-(1-y)^2}{2} \,
\frac{Q^2 - \hat{s}}{\hat{s} + Q^2} \,
\frac{\hat{u}^2 + \hat{t}^2}{\hat{u} \, \hat{t}} \; .
\end{eqnarray}
%%%%%

\vspace*{6mm}
\noindent $\bullet \:\: \gamma^\ast \, g \rightarrow Q+\bar{Q}$
(heavy quarks):

For the PGF diagram -- 
massive case ($m_Q \neq 0$, \QQ~= \cq, \bq)
the unpolarized cross section is:
\begin{eqnarray}
{\rm d} {\hat \sigma}_0 & = &
\frac{\alpha_s}{4\pi} \frac{1}{(\hat{s} + Q^2)^2} \,
\left\{ \frac{1+(1-y)^2}{2} \, \left[ \frac{Q^4 + \hat{s}^2}{(\hat{s} + Q^2)^2}
\, \frac{\tilde{u}^2 + \tilde{t}^2}{\tilde{u} \, \tilde{t}} +
\frac{ 2 \, m_Q^2}{\tilde{u} \, \tilde{t}}
\left( 2 \, (\hat{s} - Q^2) + \frac{Q^2 (\hat{s} + Q^2)^2}
 {\tilde{u} \, \tilde{t}}\right) \right. \right. \nonumber \\
& & \left. \left. - \frac{4 \, m_Q^4 (\hat{s} + Q^2)^2}
{\tilde{u}^2 \tilde{t}^2} \right] + \, (1-y) \, 8 \, Q^2
\left[ \frac{\hat{s}}{(\hat{s} + Q^2)^2} - 
\frac{m_Q^2}{\tilde{u} \, \tilde{t}}\right]\right\} \; ,
\label{mfus3}
\end{eqnarray}
%%%%%
and the polarized one is:
%%%%%
\begin{eqnarray}
{\rm d} \Delta {\hat \sigma}_0 & = &
\frac{\alpha_s}{4\pi} \frac{1}{(\hat{s} + Q^2)^2} \,
\frac{1-(1-y)^2}{2} \,
\left[ \frac{Q^2-\hat{s}}{\hat{s} + Q^2} +
\frac{2 \, m_Q^2 (\hat{s} + Q^2)}{\tilde{u} \, \tilde{t}}\right] \,
\frac{\tilde{u}^2 + \, \tilde{t}^2}{\tilde{u} \, \tilde{t}} \; ,
\label{mfus4}
\end{eqnarray}
where $\tilde{t} = m_Q^2-\hat{t}$ and $\tilde{u} = m_Q^2-\hat{u}$.

\end{document}